\documentclass[aps,prl,reprint,groupedaddress,superscriptaddress,floatfix]{revtex4-2}

\usepackage[T1]{fontenc}
\usepackage{newtxtext}
\usepackage{newtxmath}
\usepackage{bm}

\usepackage{titlesec}

\usepackage{dsfont}

\usepackage{mathtools}
\usepackage{bbold}
\usepackage{physics}

\usepackage{xcolor}
\usepackage{soul}

\usepackage{graphicx}
\usepackage[colorlinks,citecolor=blue,linkcolor=blue,urlcolor=blue]{hyperref}
\usepackage[all]{hypcap}

\usepackage{enumitem}

\usepackage{tabularx}
\usepackage{makecell}

\graphicspath{{img/}}

\usepackage{cleveref}
\usepackage{etoolbox}

\makeatletter
\let\addcontentslineOriginal\addcontentsline
\let\addcontentsline\@gobblethree
\makeatother

\begin{document}
\title{Transport Information Beyond Strong Eigenstate Thermalization}

\author{Avadhut V. Purohit}
\email{avdhoot.purohit@gmail.com}
\affiliation{Department of Physics, Visvesvaraya National Institute of Technology, Nagpur 440010, India}
\author{Harshit Sharma}
\email{harshitsharma2796@gmail.com}
\affiliation{Department of Physics, Visvesvaraya National Institute of Technology, Nagpur 440010, India}
\author{Udaysinh T. Bhosale}
\email{udaysinhbhosale@phy.vnit.ac.in}
\affiliation{Department of Physics, Visvesvaraya National Institute of Technology, Nagpur 440010, India}
\date{\today}

\begin{abstract}
    Strong eigenstate thermalization (ETH) in operationally mean-ergodic, non-chaotic systems provides an opportunity to investigate whether Floquet eigenstates retain information about classical transport beyond their leading thermodynamic behavior. We address this question using the double kicked top (DKT), which exhibits nonmixing transport and dynamics consistent with weak mixing transport under variation of its control parameters. We show that the coefficient $m$ of the leading finite-size correction to the normalized eigenstate entanglement entropy (EE) $s$ tracks the restructuring between these transport regimes even when the thermodynamic-limit entropy $c$ becomes nearly insensitive to them. This coefficient $m$ is positive for nonmixing transport and negative when dynamics consistent with weak mixing transport dominates. The coefficient's sign reversal in the coexistence regime remains stable as the semiclassical resolution increases. Our results demonstrate that finite-size eigenstate structure can encode signatures of classical transport that are invisible in the thermodynamic-limit entropy, providing an eigenstate-based diagnostic of dynamical organization beyond the leading thermodynamic behavior associated with strong ETH.
\end{abstract}

\maketitle
\textit{Introduction ---}
Understanding how classical transport structures manifest in quantum systems remains a central problem in quantum statistical mechanics~\cite{prosen2002stability, prosen2003theory, kaplan1998weak, aravinda2021dual, purohit2026strong, prosen1999ergodic}. Classically, nonmixing and mixing dynamics exhibit fundamentally different transport behavior: nonmixing transport retains long-time dynamical memory, whereas mixing transport progressively loses correlations~\cite{petersen1995ergodic, cornfeld2012ergodic, singh2024ergodic, aravinda2021dual}. Whether this distinction survives in quantum systems remains unclear.

ETH provides a successful framework for understanding thermalization in isolated quantum systems~\cite{deutsch1991quantum, deutsch2018eigenstate, srednicki1994chaos}. A recent study has shown that both ergodic and operationally mean-ergodic classical dynamics can produce strong ETH-like behavior in the quantum regime~\cite{purohit2025study}. Consequently, thermodynamic indicators alone need not distinguish quantum systems with fundamentally different underlying classical transport structures.~\cite{deutsch2018eigenstate, hallam2019lyapunov, d2016quantum}. This raises a natural question: can signatures of nonmixing and mixing transport persist in quantum eigenstates beyond leading-order thermal behavior?

To address this question, we study the DKT, which exhibits nonergodic, operationally mean-ergodic (but non-chaotic), and dynamics consistent with weak mixing regimes~\cite{haake1987classical,haake1991quantum,purohit2025study,purohit2026strong}. These regimes depend on two parameters, $k_r$ and $k_\theta$. The distinct dynamical regimes in this model therefore provide controlled settings for investigating how changes in classical transport organization are reflected in quantum eigenstates. We characterize the classical dynamics through long-time correlations and analyze the corresponding quantum structure using finite-size scaling of the EE~\cite{patil2023average}.

The EE provides a natural probe of this structure because it characterizes the bipartite organization of the entire eigenstate rather than the expectation value of a particular observable. For thermal eigenstates, the leading term in EE approaches the nearly maximal value expected for random states, with the Page result providing the corresponding benchmark~\cite{page1993average,vidmar2017entanglement}. This leading term in EE can therefore become insensitive to finer distinctions in the underlying dynamics~\cite{vidmar2017entanglement,murthy2019structure}. Finite-size corrections to this thermodynamic-limit EE, however, can reflect nonrandom structure in eigenstates~\cite{vidmar2017entanglement,murthy2019structure}. We therefore examine the finite-size scaling of EE as a probe of dynamical organization in the DKT, focusing on the leading correction to its normalized form. This provides a route to test whether information about classical transport remains encoded in Floquet eigenstates after the thermodynamic-limit EE has become insensitive to the transport regime.

We find that the coefficient of the leading finite-size correction to the normalized EE changes sign across the transport crossover and remains stable with increasing semiclassical resolution. Its strongest variation occurs where nonmixing and weak mixing-like transport coexist through intermittency. Extracted from asymptotic Floquet eigenstates rather than short-time wavepacket evolution, this coefficient establishes an eigenstate-based correspondence between classical transport organization and quantum eigenstate structure. The sign reversal is not determined by phase space delocalization alone. The Husimi participation ratio (HPR) shows that mean-ergodic nonmixing transport can already produce highly delocalized eigenstates similar to dynamics consistent with weak mixing transport. The coefficient becomes negative only when this delocalization becomes increasingly uniform, coinciding with dynamics consistent with weak mixing transport~\cite{pilatowsky2021ubiquitous,wobst2002ballistic,romera2012husimi,calixto2012husimi}.

\textit{Mixing and nonmixing transport ---}
Recent studies have shown operationally mean-ergodic behavior for large $k_\theta$ and strongly chaotic dynamics for sufficiently large $k_r$~\cite{purohit2025study, purohit2026strong}. To characterize the transport properties of these two regimes, we analyze long-time correlations of functions as follows~\cite{petersen1995ergodic,cornfeld2012ergodic,singh2024ergodic}:
\begin{align}
    C(\tau) &= \frac{1}{\tau}\sum_{t=1}^{\tau}| \langle f(\hat{T}^t x) g(x) \rangle -  \langle f(x) \rangle \langle g(x) \rangle | \quad \text{and} \notag\\
    \displaystyle \langle f(x) \rangle &= \lim_{n\to\infty} \frac{1}{n} \sum_{i=1}^{n} f(x_i).
\end{align}
Here, the evolution of the classical map is discrete, $x_{n+1}=\hat{T}x_n$, with $x$ denoting a point on the unit sphere and $\langle f(x)\rangle$ representing the long-time trajectory average. We use the Cartesian coordinates $f(x)=X$ and $g(x)=Z$ as a representative pair of observables in this paper. The corresponding results for all nine ordered combinations $f,g\in{X,Y,Z}$ are given in the supplementary material. We then define the long-time correlation amplitude as follows:
\begin{align}
    C_{\infty}=\lim_{\tau\to\infty}C(\tau).
\end{align}
The signed correlations $\langle f(\hat{T}^{t}x)g(x)\rangle-\langle f(x)\rangle\langle g(x)\rangle$ decaying to zero is the standard correlation signature of strong mixing. Our quantity $C(\tau)$ instead uses the absolute value to characterize the persistence of correlation amplitude and therefore does not directly test this signed-correlation criterion. The observed decay of the signed correlations is consistent with mixing (see Sec.~SII of supplementary material), but finite-time and finite-size fluctuations on either side of zero prevent a definitive classification from such numerical data alone. Since weak mixing has likewise not been established formally, we use the terminology \textit{dynamics consistent with weak mixing transport}. Here, we take $\tau=10^5$ as a numerical approximation to the long-time limit.


The DKT dynamics in the parameter regime $k_r \leq 1$ and large $k_\theta$ exhibit initial oscillations and approach constant correlation amplitudes. This indicates nonmixing transport (see Fig.~\ref{fig:ergodicity}(a) and Fig.~S1(a) in the supplementary material). A recent study has also shown that this regime is operationally mean-ergodic but not pointwise-ergodic. It therefore cannot be weak mixing~\cite{purohit2026strong}. In contrast, for~$k_r > 3$, the correlation amplitudes decay algebraically over longer durations and approach a small noise floor of order $10^{-4}$ (see Fig.~\ref{fig:ergodicity}(b) and Fig.~S2 in the supplementary material). The distinction between the two transport regimes is robust across nine ordered combinations of Cartesian coordinates (see the supplemental material). Throughout this paper, we therefore refer to the former (i.e., the $k_r \leq 1$ regime) as a nonmixing transport regime and the latter (i.e., $k_r > 3$ regime) as dynamics consistent with weak mixing transport.
\begin{figure}
    \includegraphics[width=0.49\linewidth]{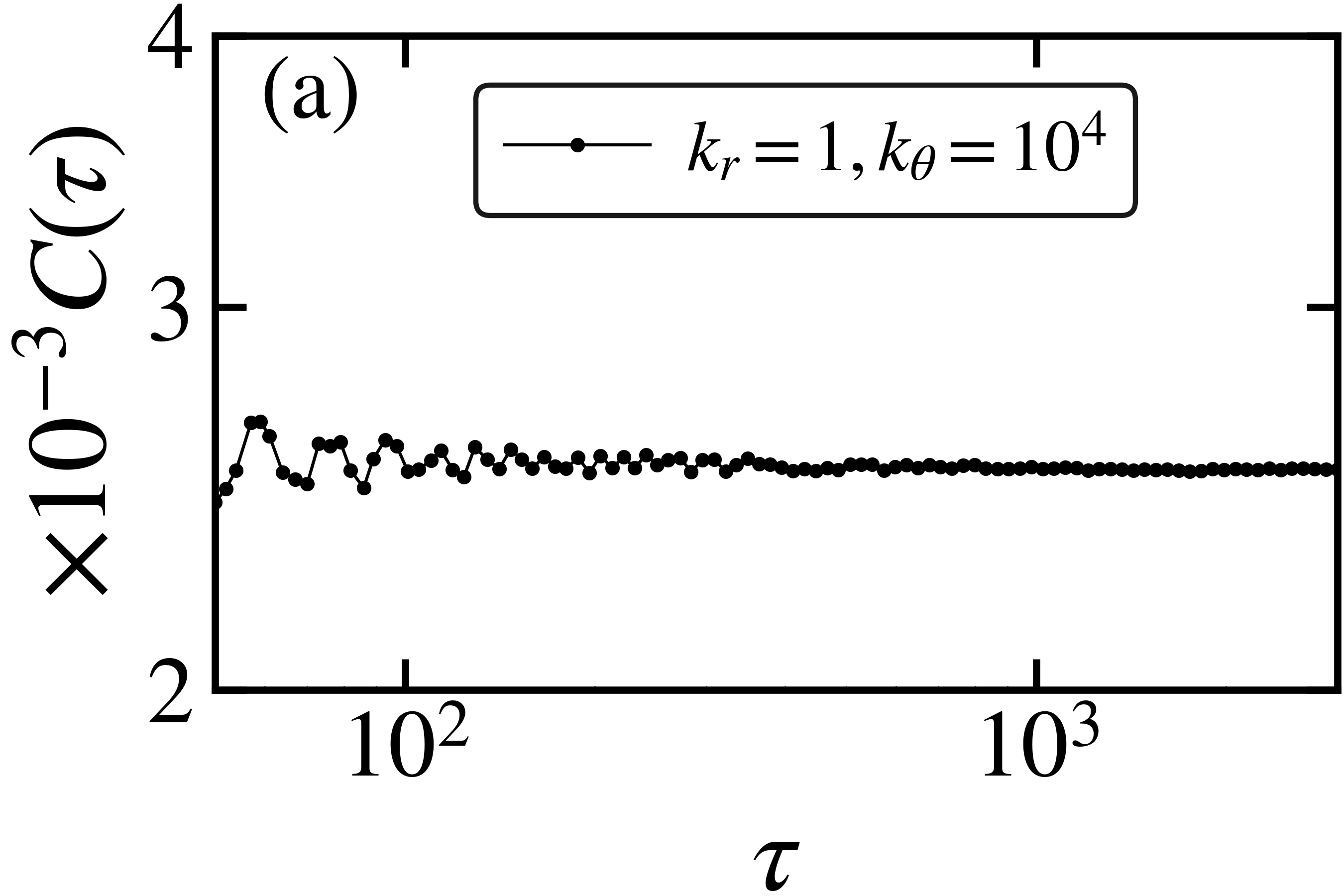}
    \includegraphics[width=0.49\linewidth]{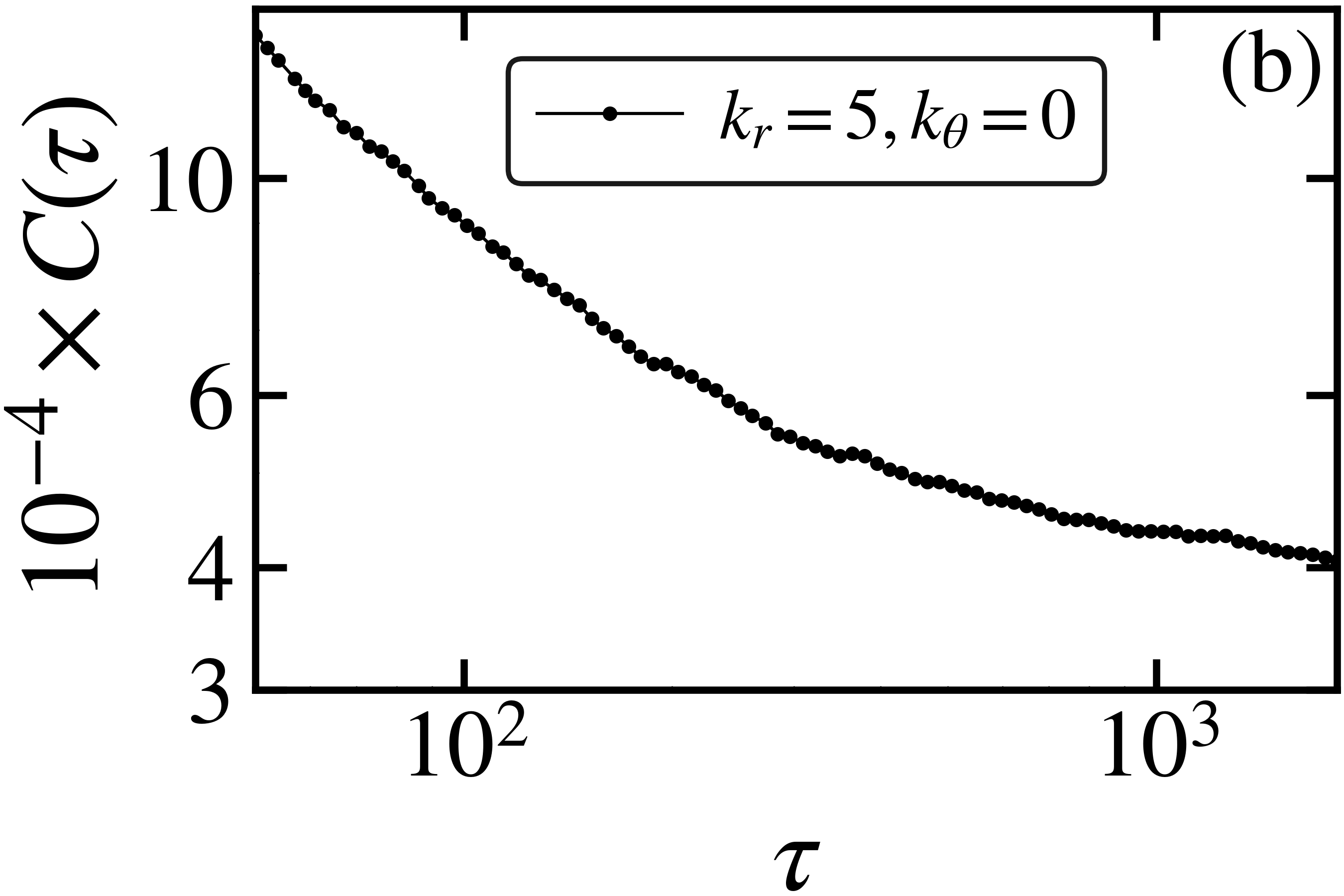}
    \caption{Correlation functions $C(\tau)$ for the DKT with initial condition $(\theta_0=1.1,\phi_0=0.3)$. (a) $(k_r=1,k_\theta=10^4)$: operationally mean-ergodic, nonmixing regime with persistent structured correlations. (b) $(k_r=5,k_\theta=0)$: regime consistent with weak mixing, characterized by algebraically decaying correlations and strongly reduced long-time correlation amplitudes.}\label{fig:ergodicity}
\end{figure}

\textit{Coexistence and intermittency ---}
The above analysis shows contrasting roles of $k_r$ and $k_\theta$ in the dynamics. While $k_r$ controls the dynamics consistent with weak mixing transport, $k_\theta$ controls operationally mean-ergodic behavior. Therefore, we fix $k_\theta = 0$ while analyzing the dynamical change from nonmixing to dynamics consistent with weak mixing transport. Our phase space analysis indicates coexistence of both transports in the parameter regime $k_r \in (1, 3)$~\cite{purohit2025study}. While trajectories in the dynamics consistent with weak mixing transport explore available regions of phase space, nonmixing trajectories retain long-time correlations. The coexistence and competition of these distinct transports give rise to intermittent dynamics (see correlations with observable $Y$ in the supplementary material)~\cite{altmann2006stickiness,geisel1987generic,zaslavsky2000hierarchical,zaslavsky2002chaos,meiss2015thirty}.

\begin{figure}
    \includegraphics[width=0.49\linewidth]{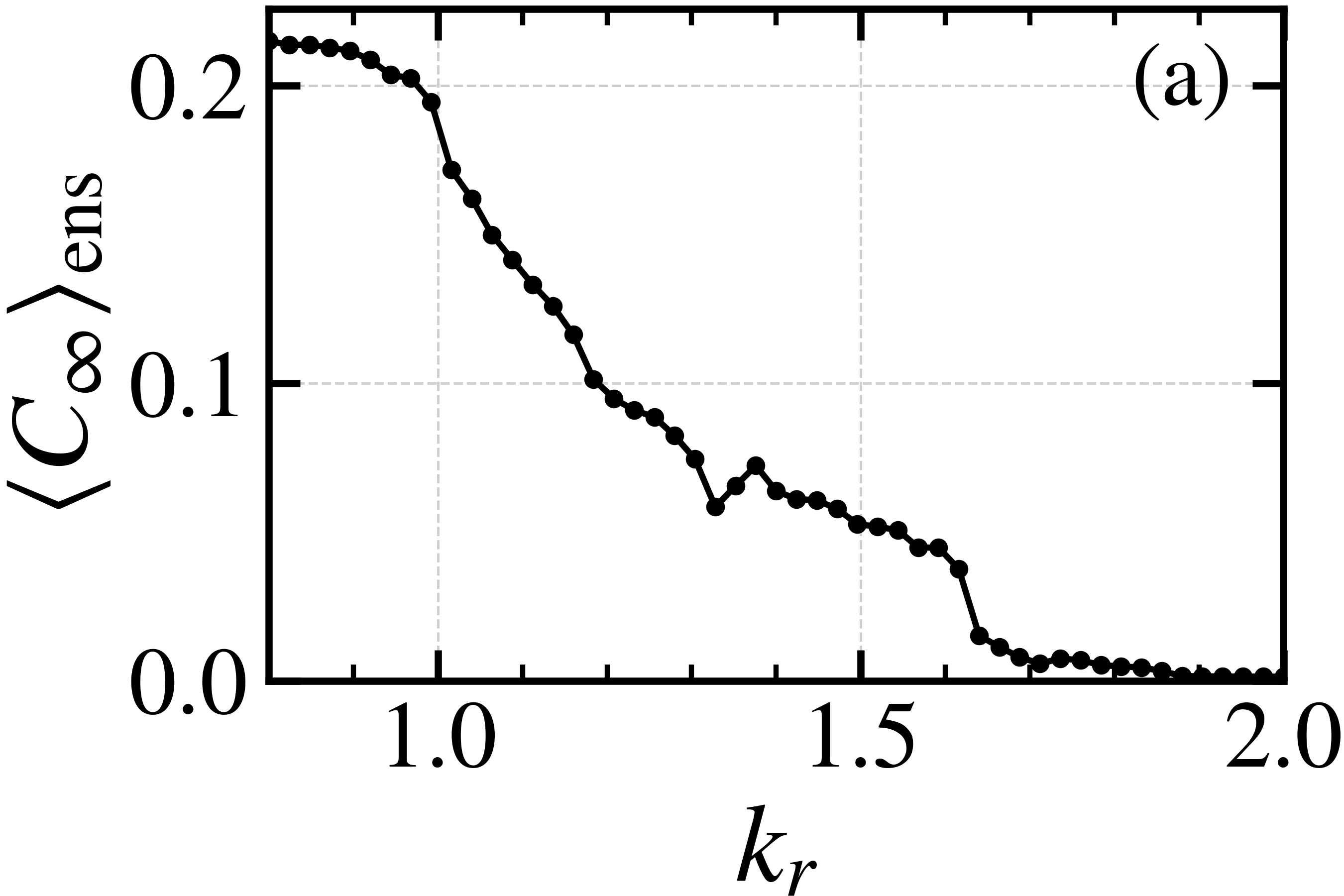}
    \includegraphics[width=0.49\linewidth]{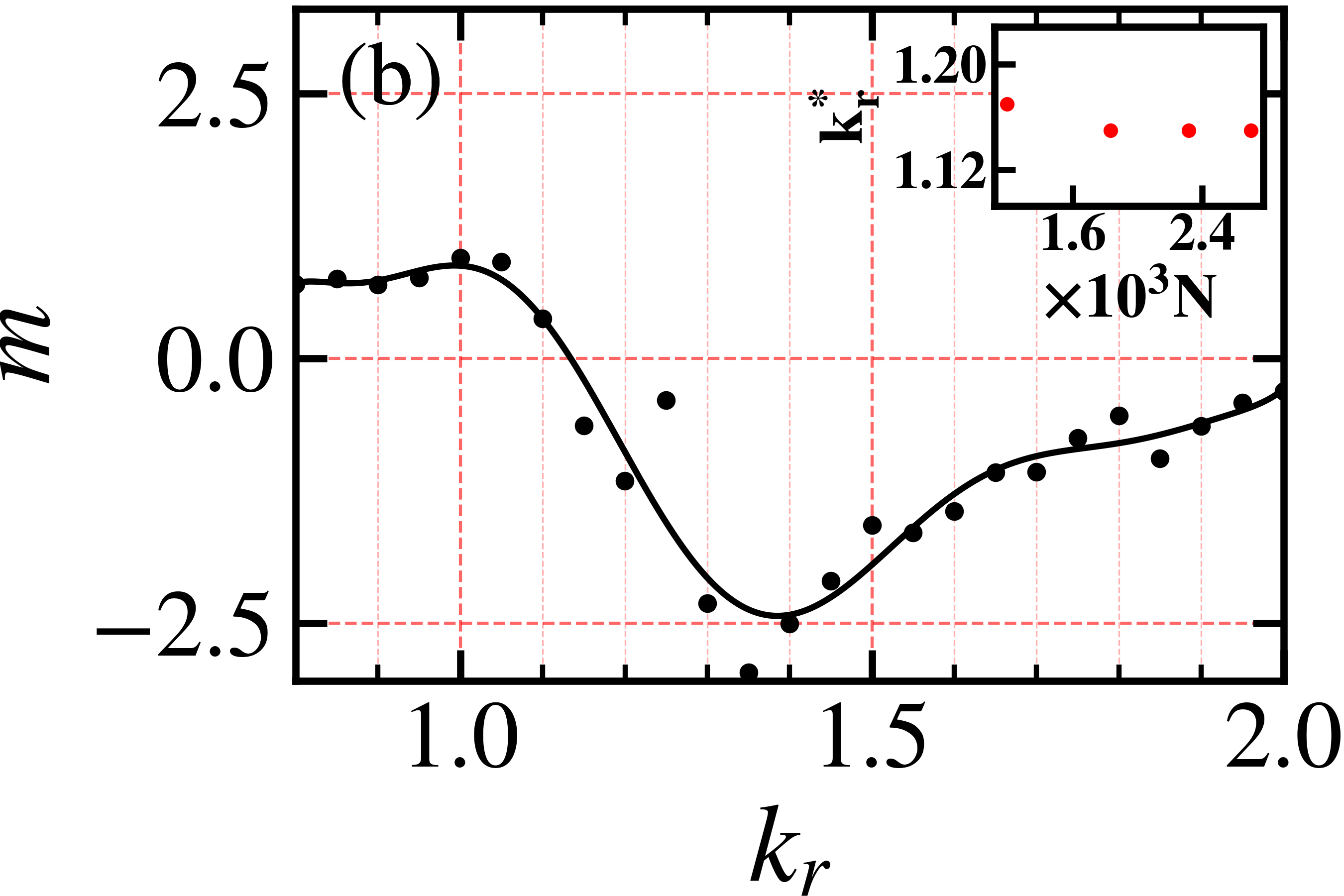}
    \caption{(a) Long-time correlation amplitudes $C_{\infty}$ averaged over the phase space as functions of $k_r$. (b) SRC obtained from finite-size scaling of the EE for system sizes in the range $N\in[3000,4000]$. For each value of $k_r$, the slope $m$ is extracted from a linear fit to 11 equispaced system sizes after resolving the parity sectors and computing the EE for a half bipartition of the system. Inset: The crossover point $k_r^{*}$, defined by $m=0$, is extracted from different finite-size windows. It remains largely stable as the system size increases. Here, $k_\theta = 0$ for both panels.}\label{fig:ainf-m}
\end{figure}

As we increase $k_r$, dynamics consistent with weak mixing transport start to emerge and domains of nonmixing transport shrink (see Fig.~1 in Ref.~\cite{purohit2025study}). The resultant transport restructuring is reflected in the long-time correlation amplitudes $C_{\infty}$ (see Fig.~\ref{fig:ainf-m}(a)). With an increase in dynamics consistent with weak mixing transport, the average phase space correlations $\langle C_\infty\rangle_{\text{ens}}$ get suppressed. Since these competing transports create intermittent trapping regions near regular structures, we can observe the persistence of long-time correlations. Therefore, the behavior of $\langle C_\infty\rangle_{\text{ens}}$ need not be monotonic with increasing $k_r$ (see correlations with observable $Y$ in the supplementary material). Eventually, for $k_r = 3$ onward, these correlations become negligible, suggesting dominance of dynamics consistent with weak mixing transport. Our analysis demonstrates that the long-time correlations provide signatures of transport restructuring.

\textit{Finite-size scaling of eigenstate entanglement ---}
A recent study has shown that both strongly chaotic and operationally mean-ergodic dynamics can exhibit strong ETH behavior~\cite{purohit2026strong}. To probe signatures of transport organization beyond leading-order thermalization, we analyze the EE of the Floquet eigenstates~\cite{page1993average, vidmar2017entanglement, garrison2018does,kliczkowski2023average}. For a half bipartition corresponding to $N$ qubits, we denote the eigenstate-averaged EE by $S$ and normalize it by using maximal entropy $S_{\max}$:
\begin{align}
    s=\frac{S}{S_{\max}}=\varepsilon S, \qquad \varepsilon=S_{\max}^{-1}.
\end{align}
Here, we use the normalized EE rather than the extensive entropy $S$. This is because the normalization separates the thermodynamic contribution from the leading finite-size correction. In the semiclassical limit, the normalized entropy approaches its thermodynamic-limit value with a finite-size correction controlled by the effective quantum coarse-graining scale $\varepsilon$. We therefore consider the leading-order scaling form:
\begin{align}
    s(N) = c + m\,\varepsilon(N) + O\!\left({\varepsilon (N)}^2\right), \label{eq:scaling}
\end{align}
where $c$ is the thermodynamic-limit value of the normalized entropy and $m$ denotes the leading finite-size correction. Higher-order terms remain numerically subleading within the explored range of system sizes.

Since $N\sim1/\hbar$ in the semiclassical limit, increasing $N$ corresponds to progressively finer semiclassical resolution of the underlying phase space structure~\cite{gutzwiller2013chaos,berry1977regular,zurek1994decoherence}. We thus define:
\begin{align}
    m=\frac{ds}{d\varepsilon}\Big|_{\varepsilon = 0},
\end{align}
and refer to it as the \emph{structure refinement coefficient} (SRC). It quantifies how the normalized EE changes as the semiclassical resolution is refined (see Fig.~\ref{fig:ainf-m}(b)). The sign of $m$ determines the direction from which the normalized entropy approaches its thermodynamic-limit value as the semiclassical resolution is increased. The zero of the SRC identifies the crossover between the two observed finite-size scaling behaviors.
 
\begin{figure}
    \includegraphics[width=0.49\linewidth]{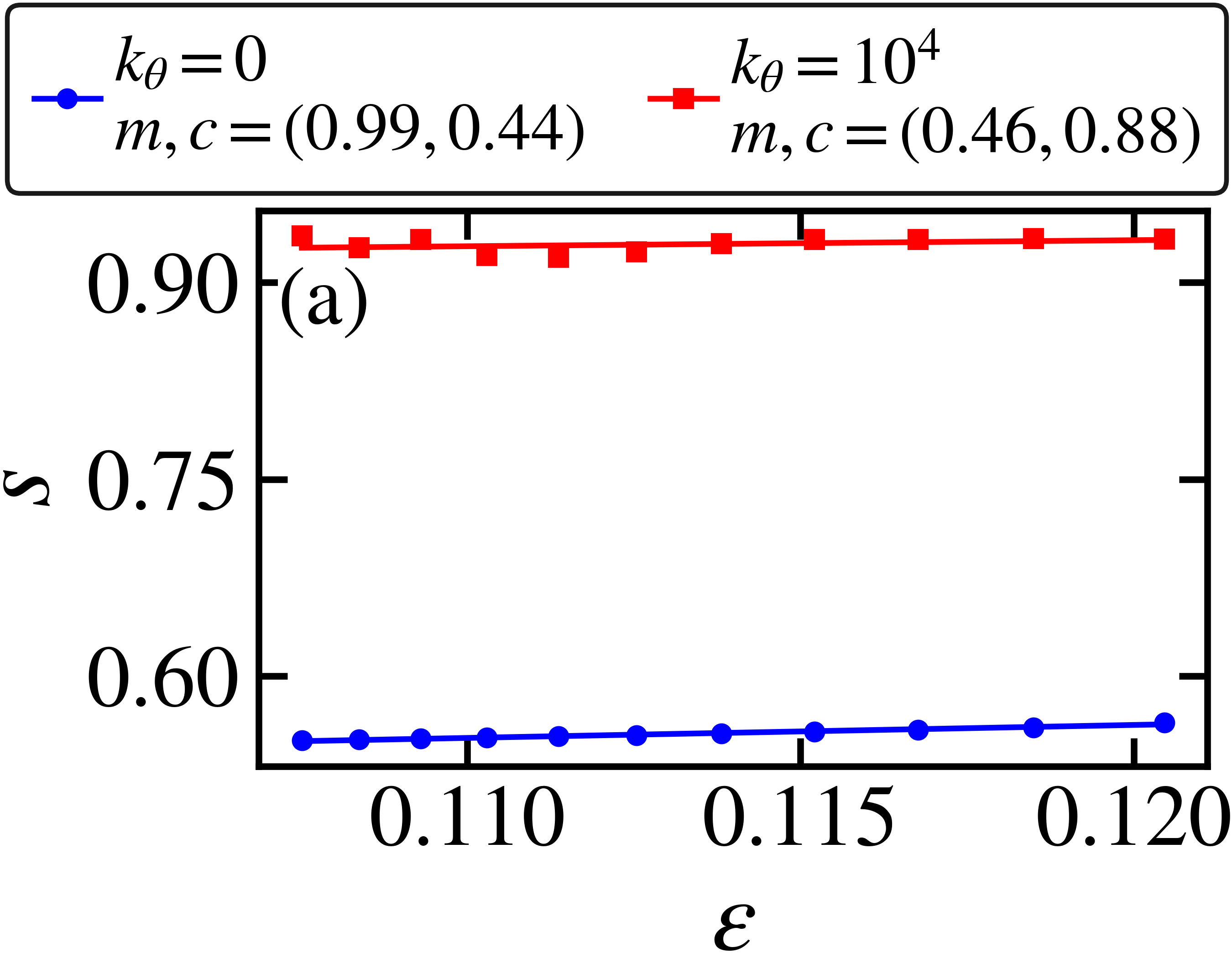}
    \includegraphics[width=0.49\linewidth]{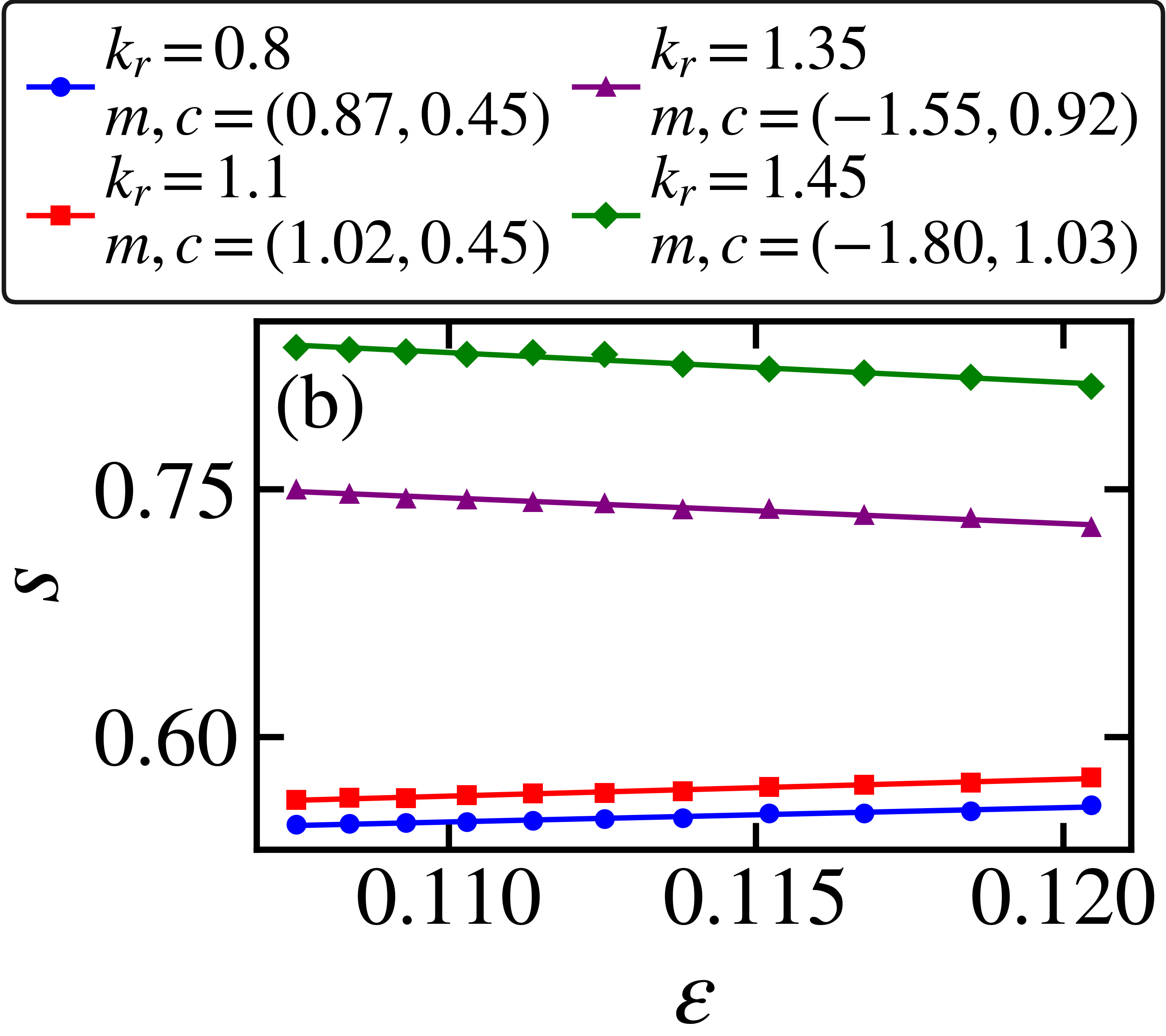}
    \caption{Normalized EE $s$ versus $\varepsilon$. (a) For fixed $k_r=1$, increasing $k_\theta$ preserves positive SRC $(m>0)$.~(b) For fixed $k_\theta=0$, increasing $k_r$ drives a sign reversal of the SRC near $k_r\approx1.20$.}\label{fig:eee-smax}
\end{figure}

The strongly chaotic (consistent with weak mixing) case $k_r\to\infty$ is known to exhibit strong ETH behavior and the case $k_\theta\to\infty$ has been recently shown to display strong ETH behavior (see Fig.~\ref{fig:eee-smax} and Ref.~\cite{purohit2026strong}). In the normalized EE, this is indicated by the thermodynamic-limit entropy $c$ approaching the Page value. Simultaneously, the SRC approaches zero from above as $k_\theta\to\infty$ and from below as $k_r\to\infty$. Since $k_\theta$ does not control dynamics consistent with weak mixing transport, we fix $k_\theta = 0$ while analyzing the SRC. As we increase $k_r > 1$, the eigenstate structure starts to reorganize. Within the narrow interval $k_r\in(1.1,1.4)$, we observe high sensitivity of the SRC (see Fig.~\ref{fig:ainf-m}(b)). Importantly, the crossover point  $k_r^{*}$ (at which we get zero SRC) remains stable against the change in system-size. This demonstrates that the observed crossover is not an artifact of the chosen finite-size window (see the inset of Fig.~\ref{fig:ainf-m}(b)).

\begin{figure}
    \includegraphics[width=0.49\linewidth]{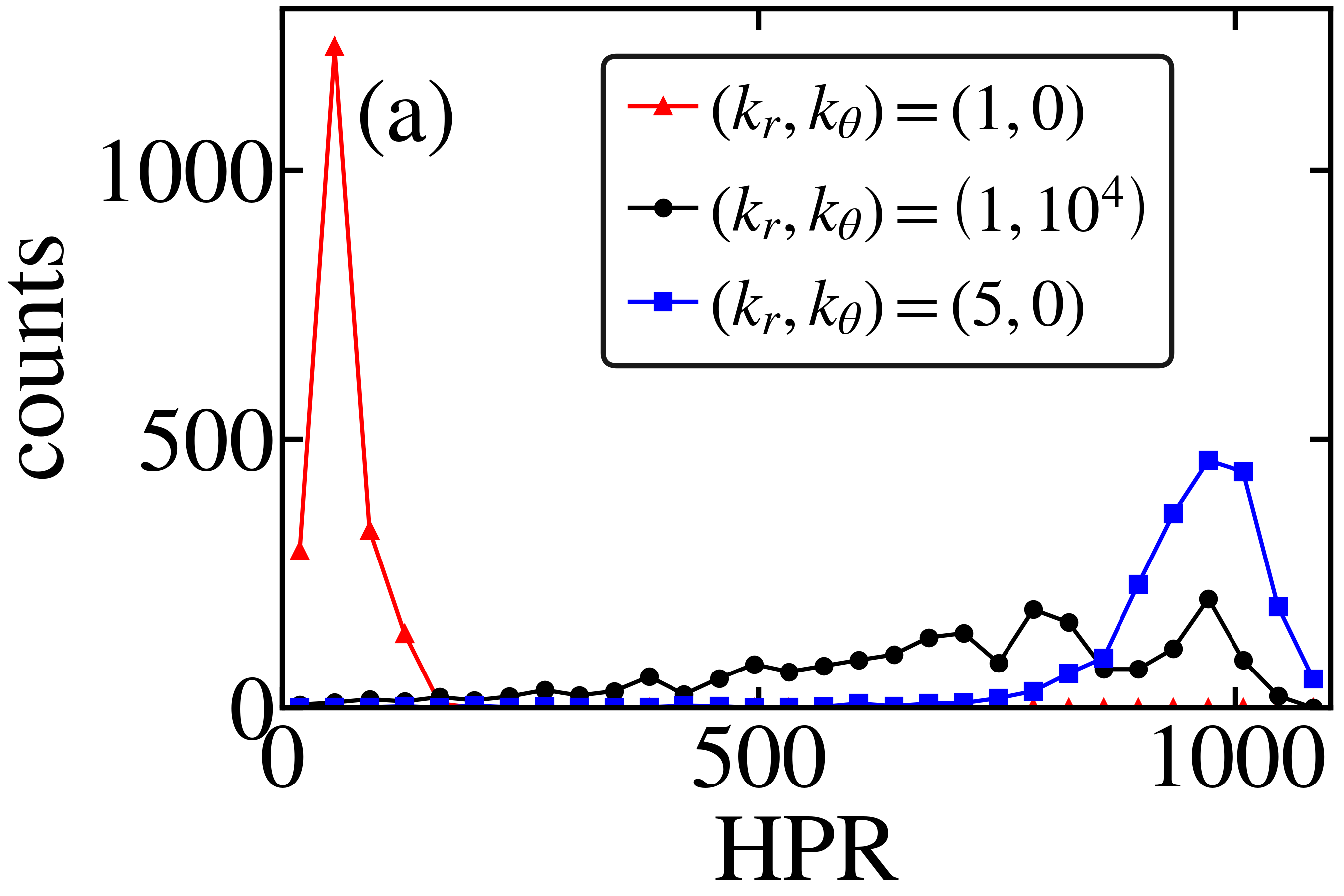}
    \includegraphics[width=0.49\linewidth]{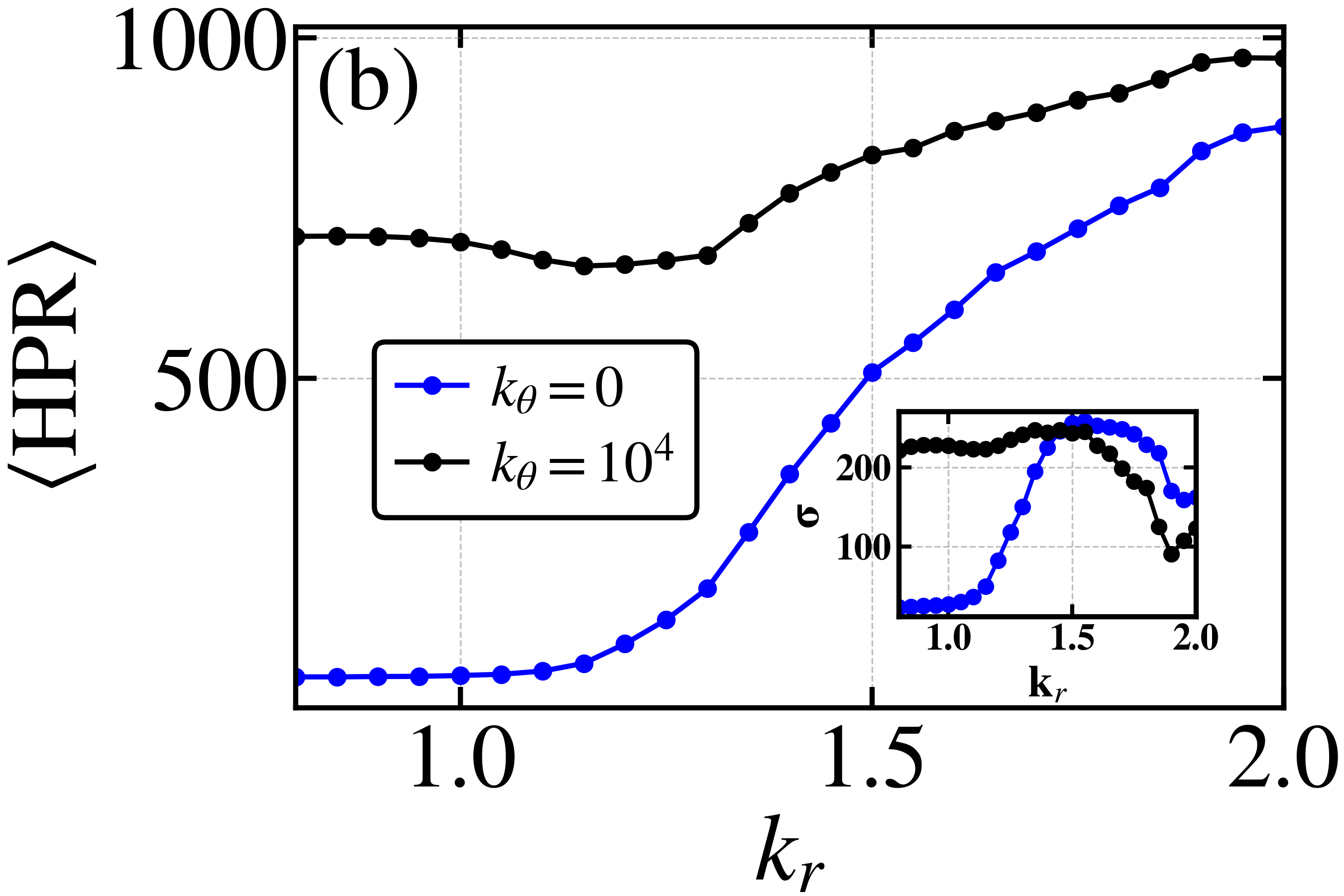}
    \caption{(a) HPR distributions for three representative cases. The HPR is concentrated at low values for $(k_r=1,\,k_\theta=0)$, spreads over a wide range for $(k_r=1,\,k_\theta=10^4)$, and is concentrated at high values for $(k_r=5,\,k_\theta=0)$.~(b) Average HPR of the Floquet eigenstates as a function of $k_r$ for spin $j=1000$. Inset: Standard deviation of the HPR over all Floquet eigenstates.}\label{fig:Husimi}
\end{figure}

\textit{Phase space delocalization ---}
We further analyze the phase space delocalization of eigenstates using the HPR to address the question of what makes the SRC negative as dynamics consistent with weak mixing transport starts to dominate~\cite{pilatowsky2021ubiquitous, wobst2002ballistic, romera2012husimi, calixto2012husimi}. The HPR is thus defined as follows:
\begin{align}
    HPR_i &= {\left(\frac{N+1}{4\pi}\int_{\Omega}Q_i^2 \; d\Omega\right)}^{-1},\\
    \langle HPR\rangle &= \frac{1}{N+1}\sum_{i=1}^{N+1}HPR_i,
\end{align}
where the Husimi function $Q_i(\theta,\phi)=|\langle\theta,\phi|\psi_i\rangle|^2$ is computed for every Floquet eigenstate $|\psi_i\rangle$. The numerical analysis shows that the nonergodic case $(k_r=1,k_\theta=0)$ exhibits low phase space delocalization, whereas the operationally mean-ergodic nonmixing case $(k_r=1,k_\theta=10^4)$ spans a broad range of HPR values despite possessing nearly Page-like entropy (see Fig.~\ref{fig:Husimi}(a)). In contrast, the dynamics consistent with weak mixing transport $(k_r=5, k_\theta=0)$ is concentrated at high HPR. These representative cases illustrate the distinct delocalization characteristics associated with the corresponding transports. 

For $k_r<1$ with $k_\theta=10^4$, the average HPR is high while the SRC remains positive (see Fig.~\ref{fig:eee-smax}(a)). This shows that a high degree of phase space delocalization alone is insufficient to produce negative SRC. The numerical results suggest that the SRC reverses sign only when the delocalization is high and substantially uniform across the phase space. This can be seen from the decreasing standard deviation $\sigma$ in the regime $k_r > 1.6$, where dynamics consistent with weak mixing starts to dominate (see the inset of Fig.~\ref{fig:Husimi}(b) and compare it with Fig.~\ref{fig:ainf-m}(a)).

\textit{Classical-quantum correspondence ---}
The long-time correlation amplitude $C_{\infty}$ quantifies the persistence of correlations in classical dynamics and measures the extent to which phase-space transport retains dynamical memory at long times. As the dynamics becomes consistent with weak mixing, classical correlations decay and the influence of initial conditions becomes progressively less persistent, resulting in a loss of dynamical memory. The corresponding quantum signatures are observed in the SRC (see Fig.~\ref{fig:ainf-m}). Unlike the thermodynamic-limit entropy $c$, the SRC provides information about how quantum correlations are distributed within an eigenstate. In this way, it distinguishes the nonuniform spreading associated with nonmixing transport from the more homogeneous spreading expected as the dynamics approaches a mixing-like regime.

The correspondence between $C_{\infty}$ and the SRC reflects a common restructuring of correlations associated with phase-space transport, although they characterize different manifestations of this restructuring. The correlation amplitude $C_{\infty}$ describes the long-time behavior of classical trajectories, whereas the SRC reveals the finite-size structure of quantum correlations in individual eigenstates. From the quantum perspective, increasing entanglement redistributes information from individual subsystems into nonlocal correlations, making it less locally accessible. In this sense, the decay of classical correlations and the redistribution of quantum correlations provide complementary signatures of dynamical memory loss and information delocalization. The SRC therefore does not provide a direct estimator of $C_{\infty}$. Rather, their correspondence is mediated by the eigenstate structure underlying thermalization.

In the spirit of ETH, thermalizing dynamics can be understood in terms of eigenstates whose local properties approach their thermal counterparts~\cite{srednicki1994chaos, deutsch1991quantum, purohit2026strong}. This leads to extensive and increasingly homogeneous entanglement as the Hilbert-space dimension grows. Conversely, persistent classical structures associated with nonmixing can leave nonuniform structure in the corresponding quantum eigenstates. From this perspective, the SRC provides a quantum eigenstate diagnostic of information organization that is reflected classically in the persistence and redistribution of dynamical correlations.

\textit{Summary and discussion ---}
We have shown that strong ETH behavior does not uniquely determine the underlying classical transport organization. In the DKT, operationally mean-ergodic nonmixing and dynamics consistent with weak mixing transport both approach the Page value in the thermodynamic limit. In contrast, the coefficient of the leading finite-size correction, quantified by the SRC, remains positive in the former transport and negative in the latter. The zero of the SRC therefore identifies a finite-size crossover at $k_r^{*} \simeq 1.15 \pm 0.04$, within the regime where the underlying transport undergoes restructuring. The HPR analysis provides complementary signatures of this restructuring. Highly delocalized eigenstates already occur in the nonmixing regime, showing that phase space delocalization alone does not determine the sign of SRC. Rather, the SRC captures how eigenstate correlations are reorganized as transport changes.

The SRC is obtained from Floquet eigenstates rather than wavepacket evolution. Therefore, it is not subject to the same Ehrenfest-time ($t_E \sim \lambda^{-1}\log N$) limitation associated with wavepacket-level classical--quantum correspondence in chaotic dynamics. Our results thus suggest that information about the organization of classical transport can remain encoded in quantum eigenstate structure even beyond the trajectory-based semiclassical regime.

The observed behavior of the SRC is also consistent with trends reported in other systems~\cite{swietek2024eigenstate, kumari2022eigenstate, vidmar2017entanglement, kliczkowski2023average, dalessio2014long, ponte2015periodically, leblond2019entanglement, patil2023average}. Positive leading finite-size correction $m$ have been found in several integrable models, including the XYZ, Lipkin-Meshkov-Glick, and Heisenberg models~\cite{swietek2024eigenstate, kumari2022eigenstate, patil2023average}, whereas negative corrections have been reported in strongly chaotic many-body and Floquet systems~\cite{vidmar2017entanglement, kliczkowski2023average, dalessio2014long, ponte2015periodically, leblond2019entanglement, patil2023average}. In this paper, we find that a single parameter $k_r$ controls the sign of SRC through nonmixing and mixing-like dynamics. Furthermore, the DKT continuously recovers the standard quantum kicked top in the limit $k_\theta=k_r$, where large kicking strengths are associated with strongly chaotic and mixing-like dynamics. These observations suggest that the transport dependence identified here may extend to other models beyond the DKT and QKT.

\textit{Acknowledgments --- }
The authors are grateful to the Department of Science and Technology (DST) for their generous financial support, making this research possible through sanctioned Project No. SR/FST/PSI/2017/5(C) to the Department of Physics of VNIT, Nagpur. 

\bibliographystyle{apsrev4-2}
\bibliography{ref}


\makeatletter
\let\addcontentsline\addcontentslineOriginal
\makeatother

\renewcommand{\thesection}{S\Roman{section}}
\renewcommand{\thesubsection}{\Alph{subsection}}
\renewcommand{\theequation}{S\arabic{equation}}
\renewcommand{\thefigure}{S\arabic{figure}}
\renewcommand{\thetable}{S\arabic{table}}

\setcounter{section}{0}
\setcounter{subsection}{0}
\setcounter{equation}{0}
\setcounter{figure}{0}
\setcounter{table}{0}

\renewcommand{\theHequation}{S\arabic{equation}}
\renewcommand{\theHfigure}{S\arabic{figure}}
\renewcommand{\theHtable}{S\arabic{table}}


\newcommand{\suppsection}[2][]{%
  \refstepcounter{section}%
  \phantomsection%
  \section*{\texorpdfstring{S\Roman{section}\quad #2}{S\Roman{section} #2}}%
  \addcontentsline{toc}{section}{S\Roman{section}\quad #2}%
  \ifstrempty{#1}{}{%
    \label{#1}%
  }%
}

\newcommand{\suppsubsection}[2][]{%
  \refstepcounter{subsection}%
  \phantomsection%
  \subsection*{#2}
  \addcontentsline{toc}{subsection}{\thesubsection\quad #2}%
  \ifstrempty{#1}{}{%
    \label{#1}%
  }%
}

\clearpage
\onecolumngrid

\begin{center}
    {\large\bfseries Supplementary Material for\par}
    \vspace{0.4em}
    {\large\itshape ``\textbf{Transport Information Beyond Strong Eigenstate Thermalization}''\par}

    \vspace{1.2em}

    {\normalsize
    Avadhut V. Purohit$^{1,*}$,
    Harshit Sharma$^{1,\dagger}$,
    and Udaysinh T. Bhosale$^{1,\ddagger}$
    \par}

    \vspace{0.6em}

    {\small
    $^{1}$Department of Physics, Visvesvaraya National Institute of
    Technology, Nagpur 440010, India
    \par}

    \vspace{0.5em}

    {\footnotesize
    $^{*}$avdhoot.purohit@gmail.com
    \quad
    $^{\dagger}$harshitsharma2796@gmail.com
    \quad
    $^{\ddagger}$udaysinhbhosale@phy.vnit.ac.in
    \par}

    \vspace{0.5em}
    {\small \today\par}
\end{center}

\vspace{1em}

\tableofcontents

\begin{figure}[htbp]
    \includegraphics[width=0.32\linewidth]{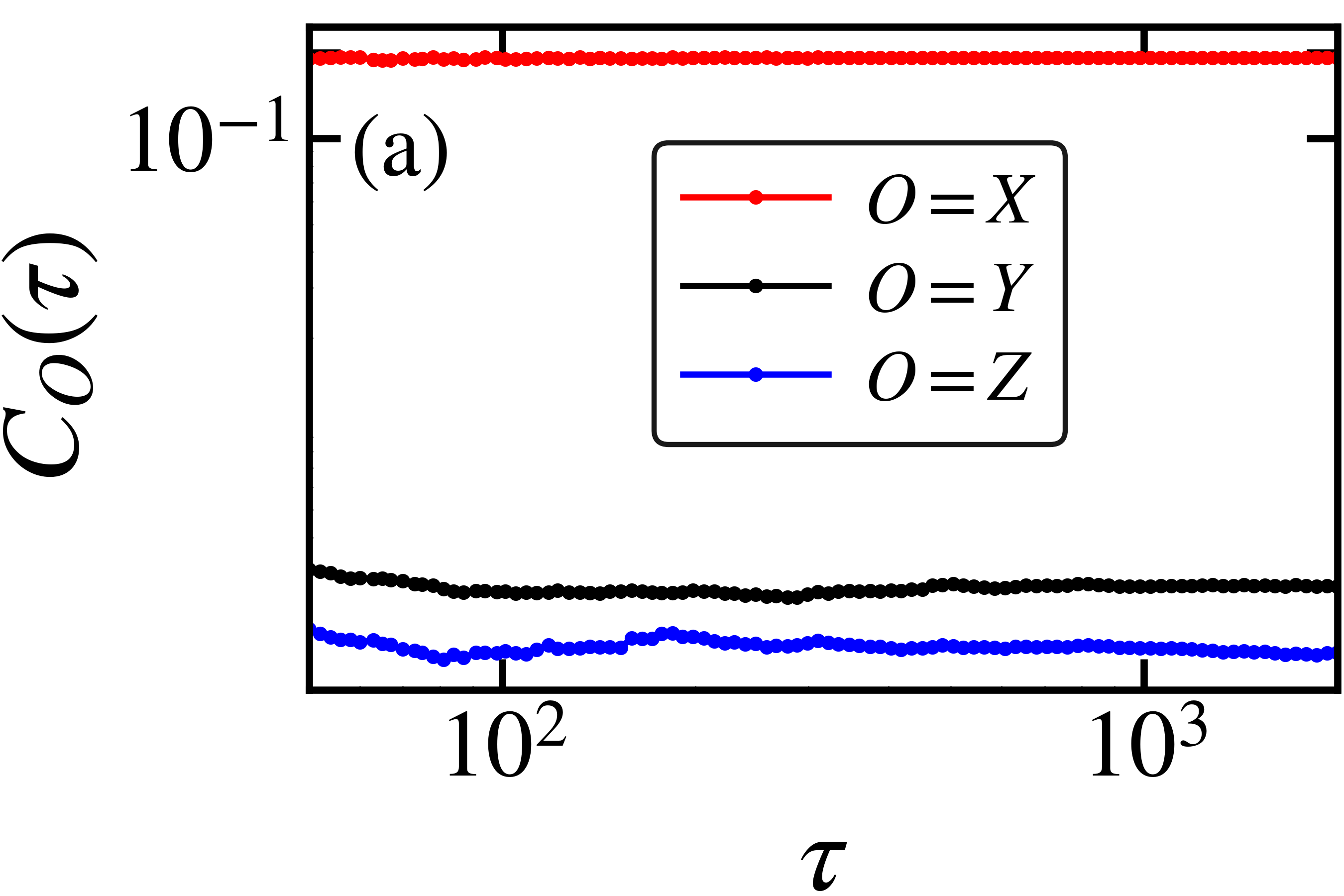}
    \includegraphics[width=0.32\linewidth]{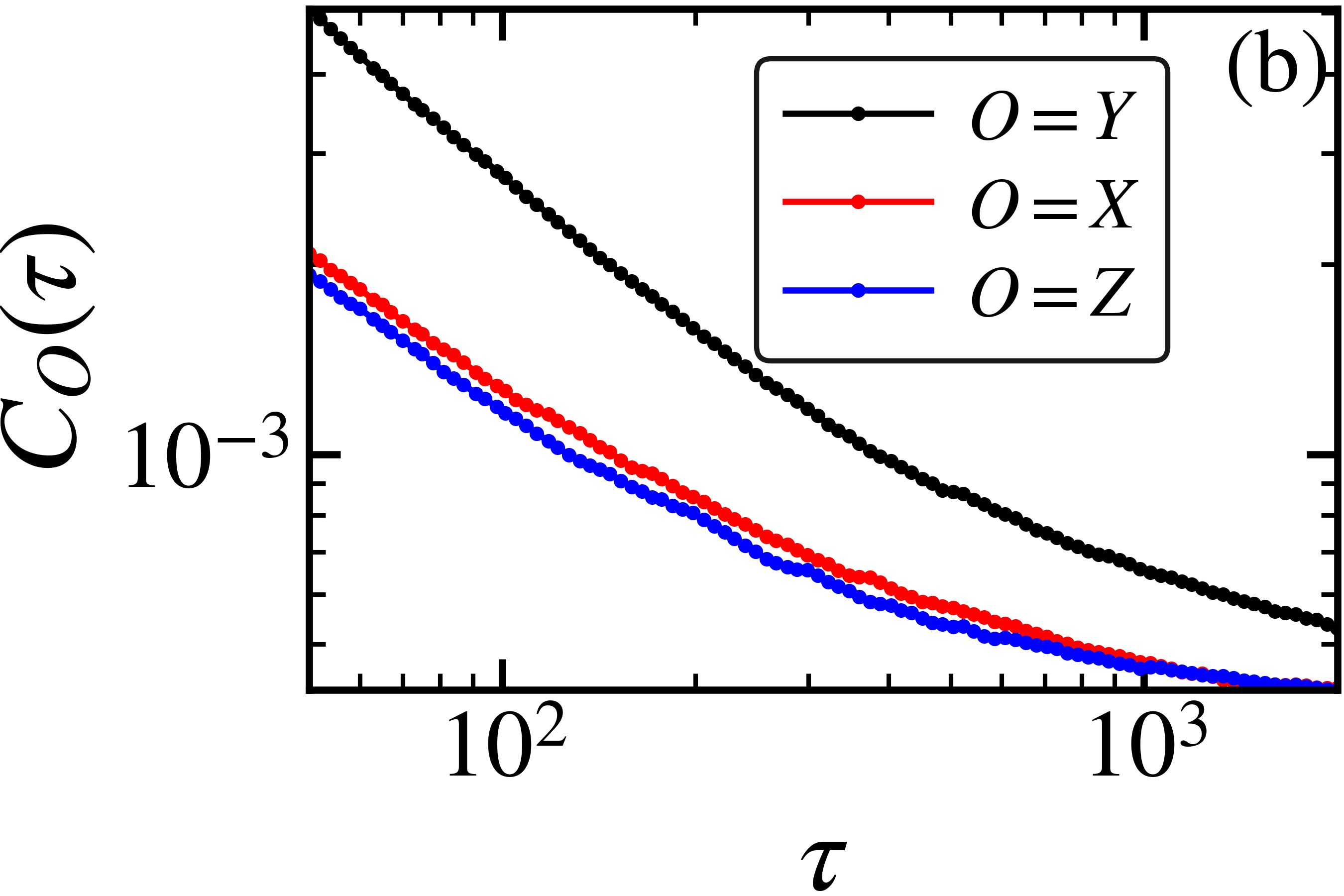}
    \includegraphics[width=0.32\linewidth]{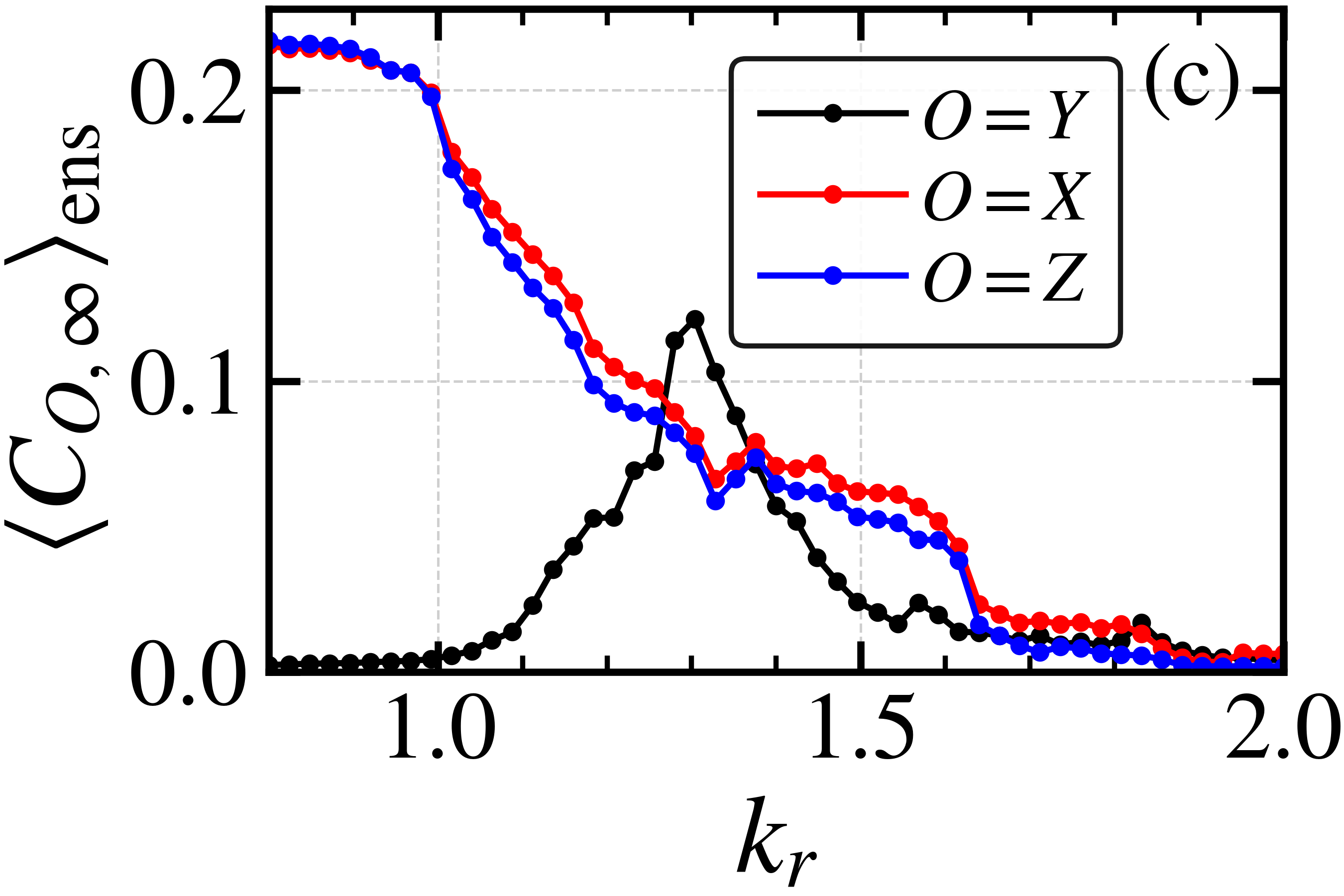}
    \caption{(a) Correlation functions $C_O(\tau)$ for the DKT with initial condition $(\theta_0=1.1,\phi_0=0.3)$ at $(k_r=1,k_\theta=10^4)$ for $O\in{X,Y,Z}$. The persistent, structured correlations at long times are characteristic of the mean-ergodic, nonmixing transport regime. (b) Correlation functions $C_O(\tau)$ for the DKT with initial condition $(\theta_0=1.1,\phi_0=0.3)$ at $(k_r=5,k_\theta=0)$ for $O\in{X,Y,Z}$. The correlations exhibit algebraic decay with time and strongly reduced long-time amplitudes, consistent with weak-mixing transport. (c) Long-time correlation amplitudes $C_{O,\infty}$, averaged over phase space, as functions of $k_r$, for the three Cartesian observables $O\in{X,Y,Z}$. Finite long-time amplitudes at low $k_r$ are associated with nonmixing transport. While their suppression at larger $k_r$ is consistent with weak-mixing transport. Here, the parameter $k_\theta = 0$.}\label{supp:fig:autocorrelation_nonmixing}
\end{figure}

\suppsection{Correlations}
The correlation functions $C_{f,g}(\tau)$ are defined in the main text. Here, we specialize to the autocorrelations by taking $f(x)=g(x)=O$, with the Cartesian coordinates $O\in{X,Y,Z}$ as observables. In the mean-ergodic, nonmixing transport~\cite{haake1987classical, haake1991quantum, purohit2025study}, the autocorrelations retain finite long-time values (see Fig.~\ref{supp:fig:autocorrelation_nonmixing}(a)). While dynamics consistent with weak-mixing transport exhibits algebraically decay of correlations (see Fig.~\ref{supp:fig:autocorrelation_nonmixing}(b)). Because of finite sampling size, these correlations settle around very small but finite floor values. These results thus support that nonmixing transport in the parameter domain $k_r \leq 1$ and dynamics consistent with weak mixing transport in the large-$k_r$ domain.

Furthermore, we numerically analyze the response of ensemble-averaged autocorrelation functions ${\langle C_{O,\infty}\rangle }_\text{ens}$ to the change in~$k_r$. Our computations show that correlations decay gradually with $k_r$ for observables $X$ and $Z$ (see Fig.~\ref{supp:fig:autocorrelation_nonmixing}(c)). In contrast, autocorrelation functions ${\langle C_{O,\infty}\rangle }_\text{ens}$ for $Y$ component start at low values in the nonmixing regime i.e., $k_r \leq 1$. This distinct behavior of the $Y$ component is because of symmetry structure of our DKT model. In the regime of coexistence, due to intermittency we observe rise of autocorrelations. This shows that autocorrelation functions are observable-dependent in the nonmixing regime and display rich dynamics in the regime of coexistence. 

\begin{figure}
    \includegraphics[width=0.32\linewidth]{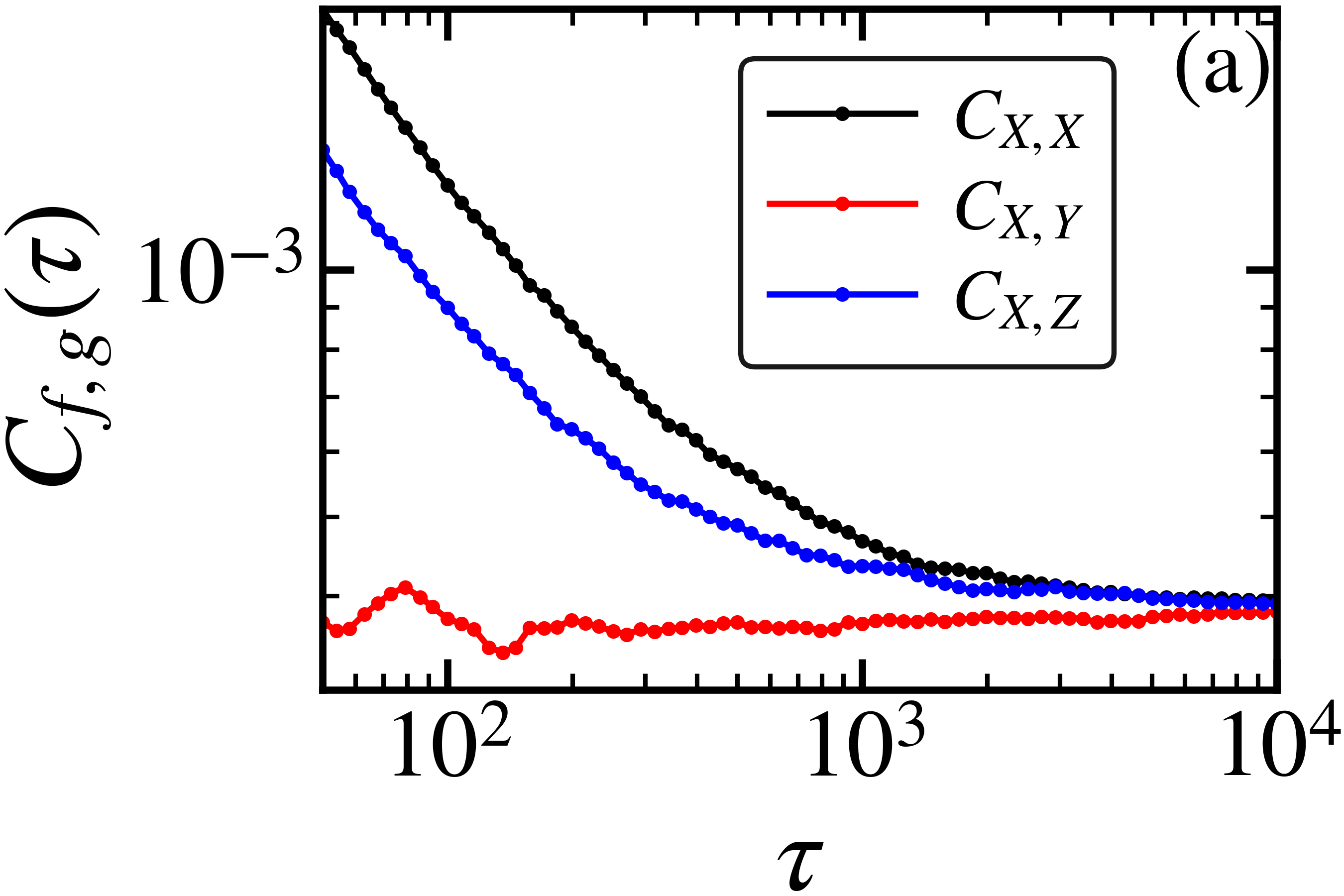}
    \includegraphics[width=0.32\linewidth]{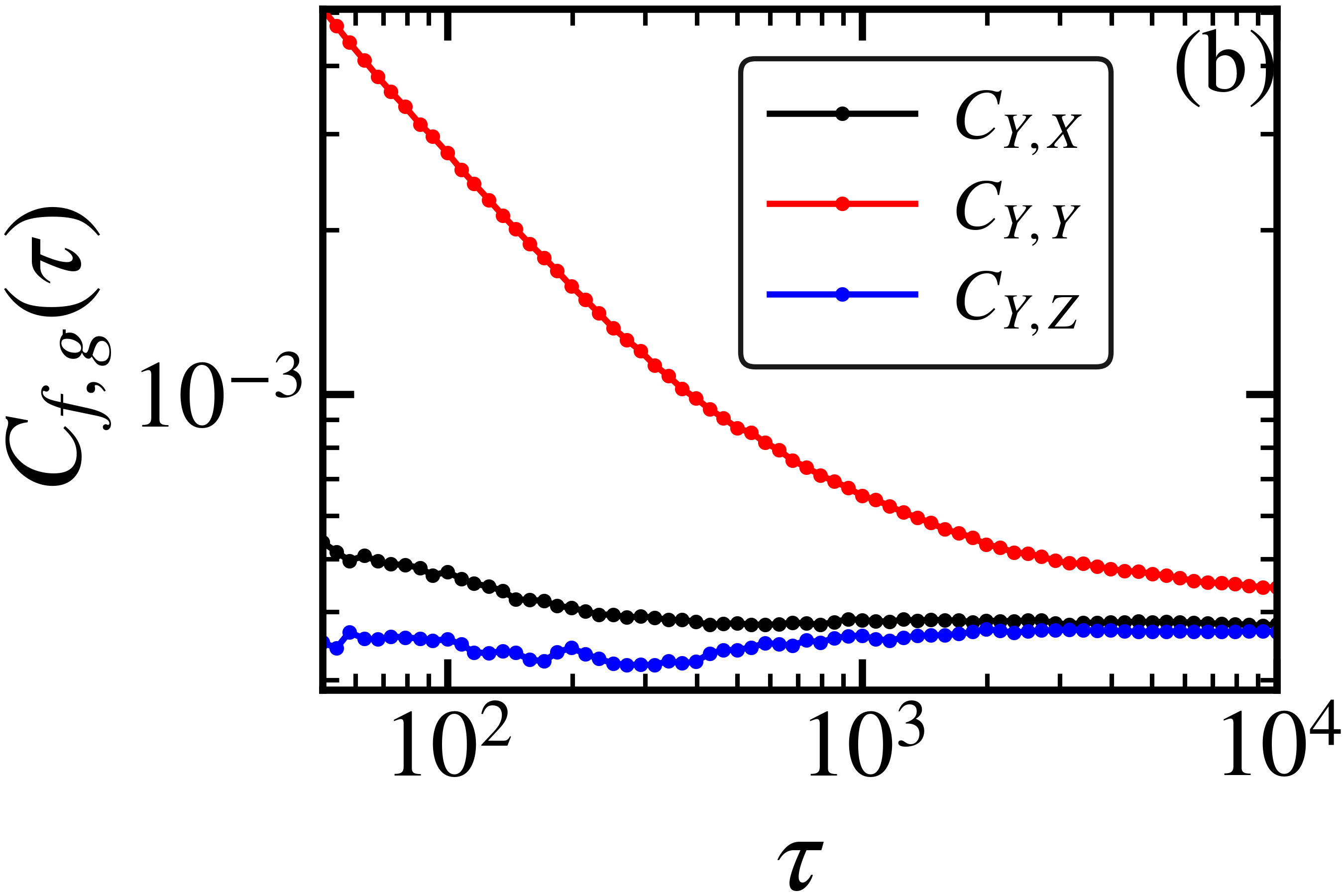}
    \includegraphics[width=0.32\linewidth]{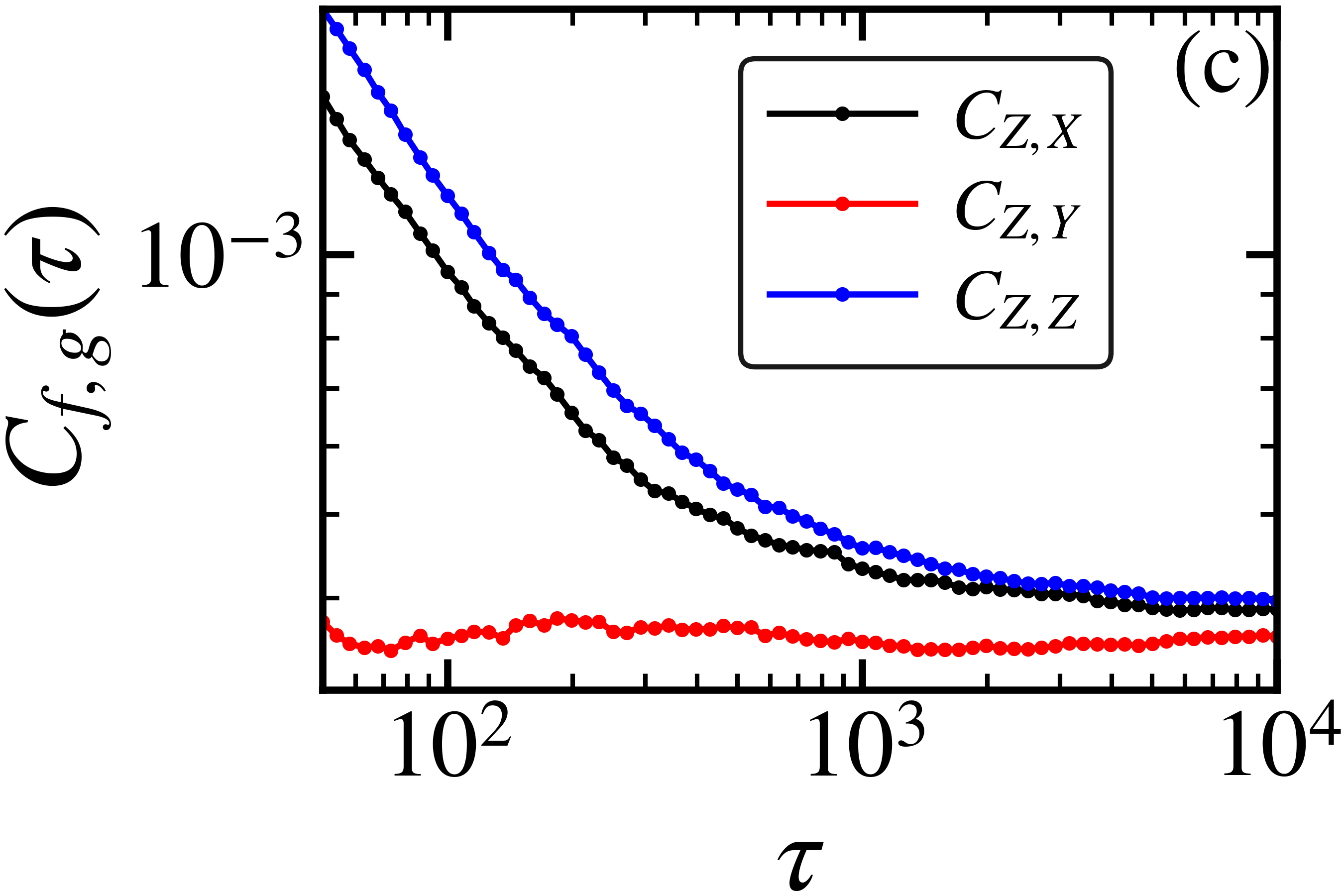}
    \caption{Time-averaged absolute two-point correlations $C_{f,g}(\tau)$ for the classical double-kicked top at $k_r=5$ and $k_\theta=0$. The nine ordered pairs of Cartesian observables are shown as $(X, X)$, $(X, Y)$, and $(X, Z)$ in (a); $(Y, X)$, $(Y, Y)$, and $(Y, Z)$ in (b); and $(Z, X)$, $(Z, Y)$, and $(Z, Z)$ in (c). The decay of the correlations toward the numerical floor is consistent with the weak-mixing-like transport identified in the main text. Since, $\langle f \rangle \sim 10^{-4}$ for Cartesian coordinates, we consider $C_{f,g}(\tau = 10^5) \sim 10^{-4}$ as operational zero.}\label{supp:fig:strong_mixing_correlations}
\end{figure}

Since dynamics in the domain $k_r \leq 1$ and $k_\theta = 10^4$ is not weak mixing~\cite{purohit2026strong}, we only analyze domain $k_r > 3$ for dynamics consistent with weak mixing transport for few other variables (see Fig.~\ref{supp:fig:strong_mixing_correlations}). Our computations show that correlation functions $C_{f,g}(\tau )$, defined in the main text, either decay to floor value or start at floor value itself (latter is due to symmetry structure of our DKT model). This shows that the suppression of long-time correlations is not restricted to a particular choice of observable. The decay of all nine correlation functions show numerical evidences for dynamics consistent with the mixing transport.

For all three Cartesian observables, we expect $\langle f \rangle$ to be zero. However, due to finite-size of our sampling we get numerical floor value of the order $10^{-4}$. Consequently, correlation functions fluctuate around this residual floor value. Higher sampling further reduces this floor value. Therefore, we consider fluctuations of the order $10^{-4}$ as operational zero.

\suppsection{Mixing-like dynamics}
In this section, we give numerical evidences for dynamics consistent with strong mixing. However, as we discussed in the main text, we have not formally established weak mixing for the regime of interest. Therefore, the following numerical analysis only gives evidences for dynamics consistent with strong mixing. 

\begin{figure}
    \centering
    \includegraphics[width=0.49\linewidth]{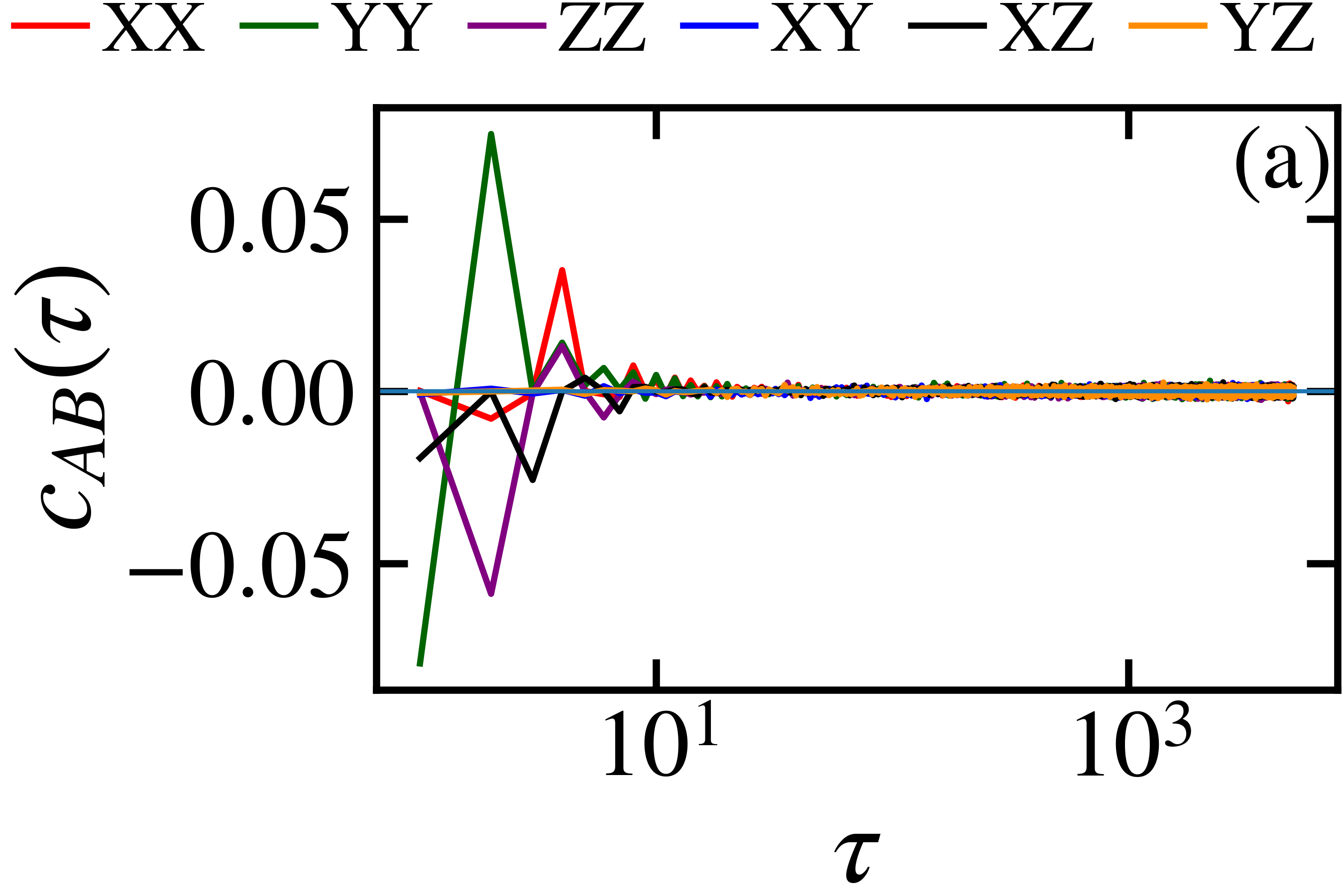}
    \includegraphics[width=0.49\linewidth]{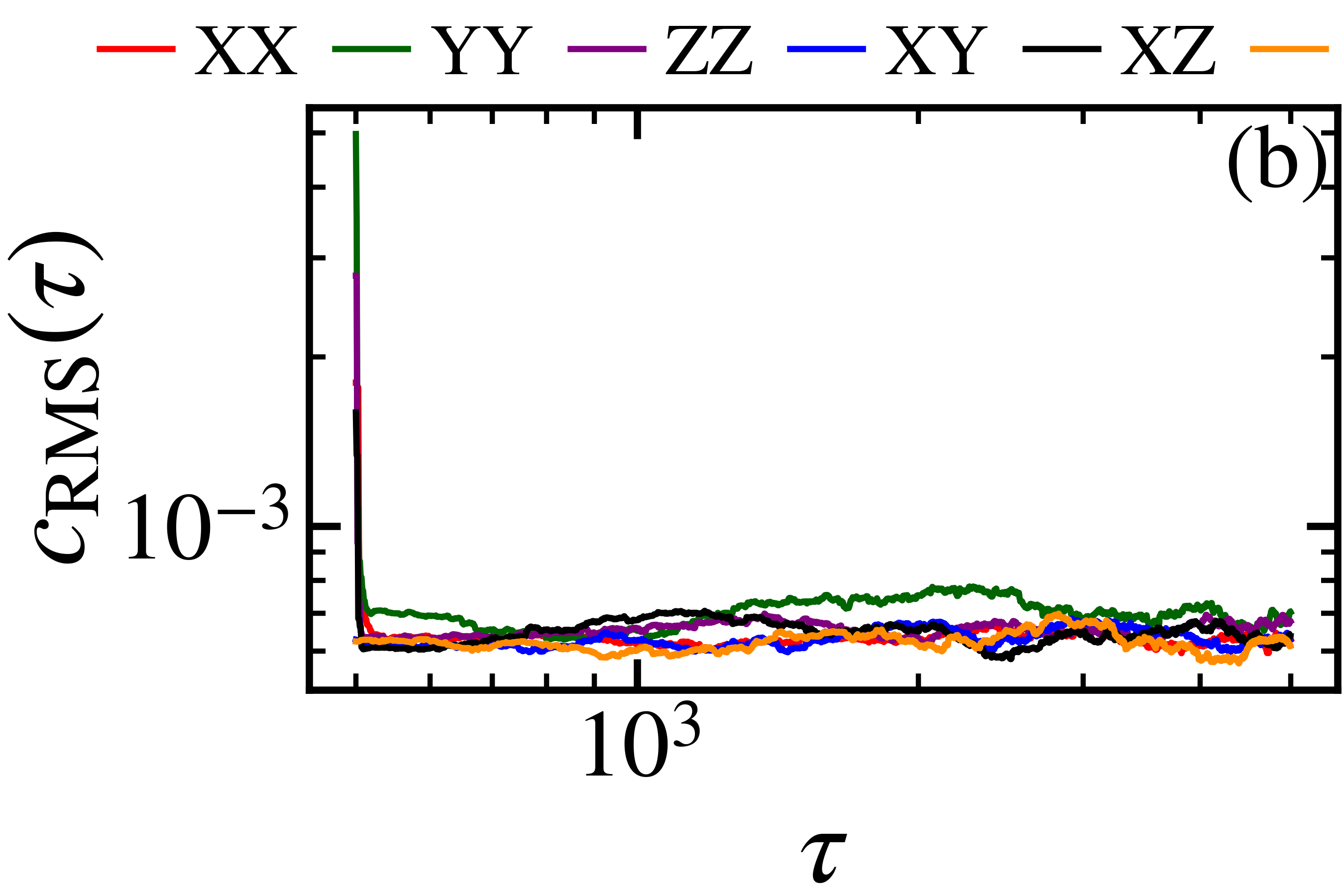}
    \includegraphics[width=0.49\linewidth]{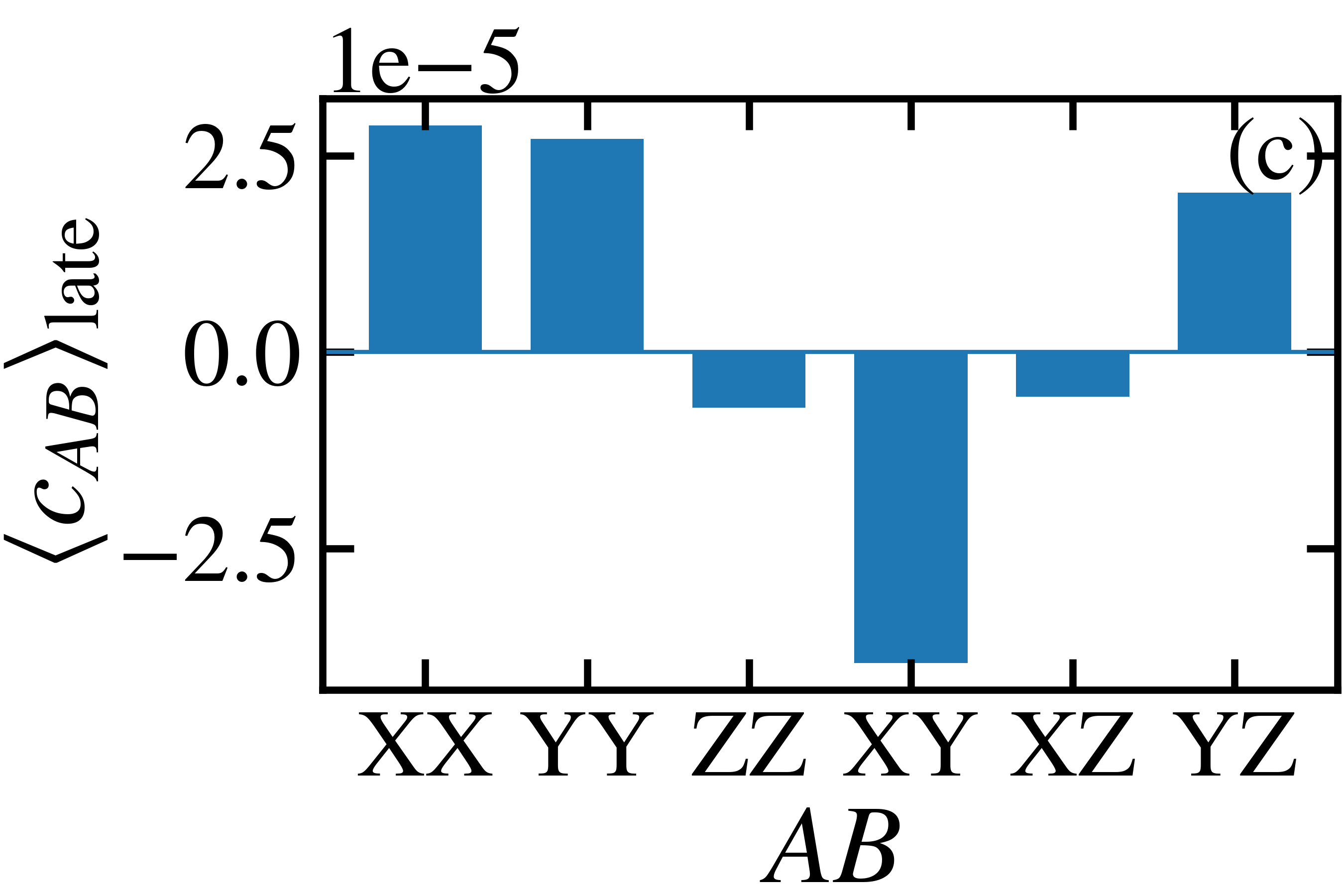}
    \includegraphics[width=0.49\linewidth]{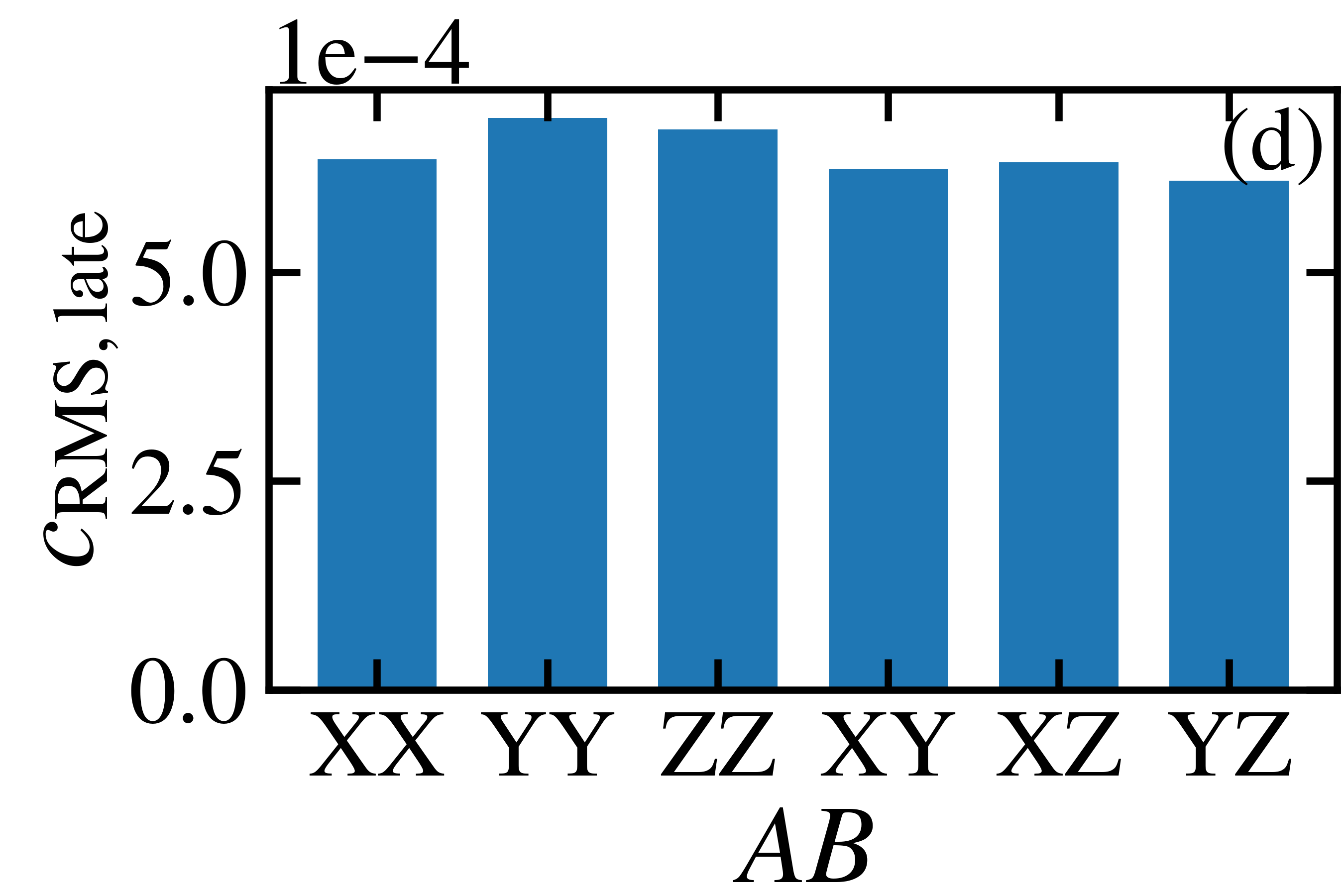}
    \caption{(a) Correlation functions $c_{AB}(\tau)$ for $k_r = 5$ and $k_\theta = 0$. (b) Sliding-window RMS correlations corresponding to $c_{AB}(\tau)$. The RMS quantifies the local amplitude of the correlation fluctuations. (c) Late-time correlations obtained from the interval $\tau_{\mathrm{LT}} \leq \tau \leq 10^5$. The late-time-averaged correlations provide a finite-time diagnostic of the residual correlations. (d) Late-time RMS correlations. They remain finite when persistent oscillations cancel upon signed averaging and therefore provide a diagnostic of residual long-time correlation amplitude.}\label{fig:supp_correlations_signed}
\end{figure}

Now, for an observable $A$ evaluated after $\tau$ iterations and an initial observable $B$, we define the following correlation function~\cite{petersen1995ergodic, cornfeld2012ergodic}:
\begin{align}
    c_{AB}(\tau) &= \left\langle \left[A(T^\tau x)-\left\langle A(T^\tau x)\right\rangle\right] \left[B(x)-\left\langle B(x)\right\rangle\right] \right\rangle , \label{eq:supp_signed_corr} \\ 
    c_{\mathrm{RMS}}(\tau;W) &= \left[ \frac{1}{W} \sum_{s\in\mathcal{W}_\tau} c_{AB}(s)^2 \right]^{1/2}, \label{eq:supp_sliding_rms} \\
    c_{\mathrm{LT}} &= \frac{1}{N_{\mathrm{LT}}} \sum_{\tau=\tau_{\mathrm{LT}}}^{\tau_{\max}} c_{AB}(\tau), \quad N_{\mathrm{LT}}=\tau_{\max}-\tau_{\mathrm{LT}}+1 \quad \text{and} \label{eq:supp_late_corr} \\ 
    c_{\mathrm{RMS,LT}} &= \left[ \frac{1}{N_{\mathrm{LT}}} \sum_{\tau=\tau_{\mathrm{LT}}}^{\tau_{\max}} c_{AB}(\tau)^2 \right]^{1/2}. \label{eq:supp_late_rms}
\end{align}
Here, $T$ denotes the classical DKT map and $\langle\cdots\rangle$ denotes an average over the uniformly sampled phase-space ensemble. We consider a few correlations such as $c_{XX}$, $c_{YY}$, $c_{ZZ}$, $c_{XY}$, $c_{XZ}$ and $c_{YZ}$. Our numerical results show that some correction functions quickly decay and fluctuate around zero (see Fig.~\ref{fig:supp_correlations_signed}(a)). While others start around zero and fluctuate around it. This shows that dynamics in the region of interest is consistent with strong mixing.

Because of these fluctuations around zero, we additionally evaluate the sliding-window root-mean-square (RMS) correlation functions $c_{\mathrm{RMS}}(\tau;W)$ (see Fig.~\ref{fig:supp_correlations_signed}(b)). This RMS quantity is insensitive to the sign of the oscillations and provides a local measure of their remaining amplitude. The corresponding late-time signed correlations $c_{\mathrm{LT}}$ are shown in Fig.~\ref{fig:supp_correlations_signed}(c), where $\tau_{\max}=10^5$ is used as a practical approximation to the long-time regime. Finally, the late-time RMS correlation $c_{\mathrm{RMS,LT}}$ are illustrated in Fig.~\ref{fig:supp_correlations_signed}(d). These late-time RMS correlation functions fall within floor value and therefore consistent with mixing dynamics. Here, $\tau_{\mathrm{LT}}$ is of the order $10^4$. 

\end{document}